# MODELING OF THE FINAL JUMP PHASE OF A POSITIVE LIGHTNING LEADER

N I Petrov and G N Petrova

**ABSTRACT**

Behavior of the potential and electric field intensity at the front of a lightning leader channel taking into account the influence of space charges injected by the streamers is investigated theoretically. Analytical solutions of Poisson equation are found for different laws of charge density distribution in the streamer zone. Influence of the streamer zone parameters on the formation of final jump phase of a leader is investigated.

## 1. INTRODUCTION

The striking distance of a lightning to structures is assumed to be dependent of the lightning peak current in the return stroke, which in its turn is a function of a lightning leader charge. Development of a positive lightning leader takes place with a practically constant velocity, so the linear charge density does not change distinctly in the process of leader propagation. However, an intensification of electric field intensity between streamer zone front and earth surface takes place at the approaching of a leader to the ground surface owing to the effect of "image" charge. This causes an abrupt increase in the leader-streamer system velocity and current, streamer zone length and its charge. This stage of leader propagation is referred to as final jump phase. The streamer zone length at final jump phase determines the striking distance to structures, so the analysis of this phase is of practical interest.

The total charge of streamer zone and the potential of a leader head depend on the law of charge density distribution in a streamer zone. Therefore the investigation of potential and electric field distributions in the dependence on the streamer zone parameters for different laws of charge density distributions is of practical interest.

Determination of the volume charge distribution from the analysis of the system of equations describing streamer corona formation is not possible at present. Measurements of the charge density inside the streamer corona are also difficult. There are only measurements of the electric field intensity in a streamer corona and streamer zone of a leader [1-3]. Particularly, it was shown in [2] that the amplitude of the electric field intensity in a streamer corona and streamer zone of a positive leader is about 5 kV/cm. It was shown in [3] that this value is kept along the total length of streamer corona and streamer zone of a leader. This result is important so far as it allows determination of the charge density distribution from the Poisson equation. So the determination of charge density distribution may be reduced to the investigation of the behavior of potential and electric field in a discharge gap for various parameters of a streamer zone.

Many works were devoted in recent years to the modeling of propagation of positive streamer-leader system in air discharge gap [4-12].

In this paper theoretical investigations of the formation of streamer zone of positive leader in an atmosphere taking into account the influence of space charges injected by the streamers itself are carried out. It is shown that the potential and electric field distributions inside the streamer corona depend on its geometrical parameters. The volume charge distribution law in a streamer corona is found, for which the electric filed intensity is kept along the axis line. Propagation of the leader channel/streamer zone system taking into account the influence of the "image" charges are carried out. It is shown that the streamer zone length increases sharply at the final jump phase when the streamer zone front reaches the ground surface.

## 2. POTENTIAL AND ELECTRIC FIELD DISTRIBUTIONS

Geometrical scheme of volume charge distribution in a streamer corona is presented in Fig.1.

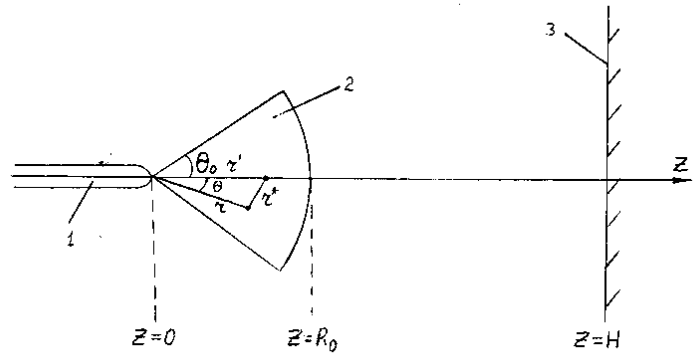

Fig. 1. Streamer-leader system in long air gap.

The potential distribution may be found from the Poisson equation:

$$\Delta\varphi = -\frac{4\pi}{\varepsilon_0}\rho, \qquad \varphi = \varphi_{el} + \varphi_v, \qquad (1)$$

where $\rho$ is the charge density, $\varepsilon_0 = 8.85 \cdot 10^{-12}$ F/m is the dielectric constant, $\varphi_{el}$ is the potential of the electrode, $\varphi_v$ is the potential distribution of volume charges.

The electric field intensity may be found from the equation $E = -\,grad\varphi$. High voltage electrode may be represented by the ellipsoid. Then the potential of electrode taking into account the influence of earth surface has the form:

$$\varphi_{el} = \frac{U_0}{arch(a/b)} \left[ arth \frac{\sqrt{a^2-b^2}}{z+a} - arth \frac{\sqrt{a^2-b^2}}{2H+a-z} \right] \qquad (2)$$

where $U_0$ is the potential of electrode, $a$ and $b$ are the half-axes of the ellipsoid.

The solution of the equation (1) for $\varphi_v$ may be represented in an integral form

$$\varphi_v = \frac{1}{4\pi\varepsilon_0} \left[ \int \frac{\rho dV}{r^*} - \int \frac{\rho dV}{R^*} \right], \qquad (3)$$

where the second term corresponds to the potential created by "image" charges, $dV = 2\pi r^2 \sin\theta d\theta dr$, $r^* = \sqrt{r^2 + r'^2 - 2rr'\cos\theta}$, $R^* = \sqrt{r^2 + (2H-r')^2 + 2r(2H-r')\cos\theta}$.

For the points on the axis line of discharge gap the analytical expressions for potential and electric field may be obtained for different charge density distribution laws.

In the case of spherical symmetry of charges distribution it is followed from the Poisson equation that the charge distribution $\rho \sim 1/r$ satisfies to the condition of electric field is kept constant inside the streamer corona. However the streamer corona represents itself the cone area, i.e. does not have the spherical symmetry (Fig.1). So, the angle on the top of cone $\Omega = 2\theta_0$ in a long laboratory gaps changes in the range of $\Omega \approx 30°$-$90°$ [2, 13, 14]. Investigation of electric field behavior in the case of cone geometry of charges distribution is of interest. Calculating the integral (3) we obtain the following expression for the potential:

$$\varphi_v = \frac{\rho_0 R_0}{2\varepsilon_0} \left\{ \frac{f(z)}{z} - \frac{f(2H-z)}{2H-z} \right\}, \qquad (4)$$

where $\rho_0$ is the charge density in the boundary of streamer corona,

$$f(z) = \frac{1}{2} A \cdot B + \frac{z^2 \sin^2\theta}{2} \ln \left| \frac{A+B}{z - z\cos\theta} \right| + \frac{z^2 \cos\theta - z^2}{2} + \frac{z - R_0}{2} A,$$

$$A = \sqrt{z^2 + R_0^2 - 2zR_0 \cos\theta_0}, \quad B = R_0 - z\cos\theta.$$

The function $f(2H-z)$ may be obtained from the function $f(z)$ by simple replacement of $z$ to $2H-z$.

The distributions of electric field on the axis line of a gap are presented in Fig. 2a for different values of the streamer corona top angle $\theta_0$.

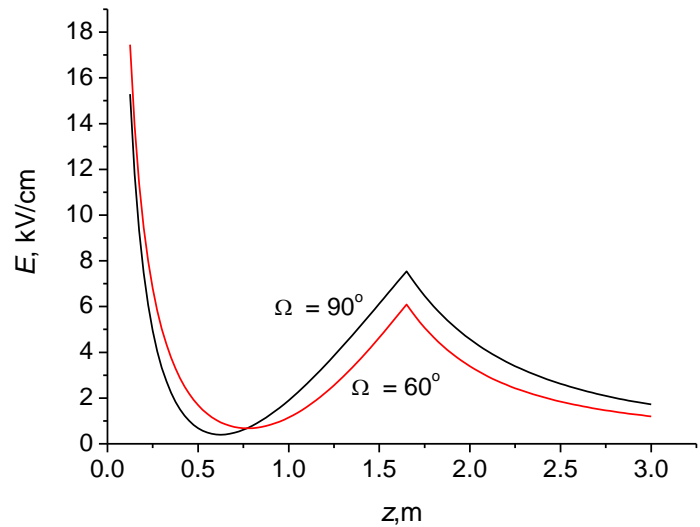

Fig. 2a. Electric field distribution in the streamer zone of a leader: $\rho \sim 1/r$.

Consider the uniform distribution of charges in a streamer corona: $\rho = \rho_0$. In Fig.2a the electric field distributions are presented for different values of angles $\theta_0$. Calculations show that in none of the angles $\theta_0$ the electric field does not keep along the streamer corona length.

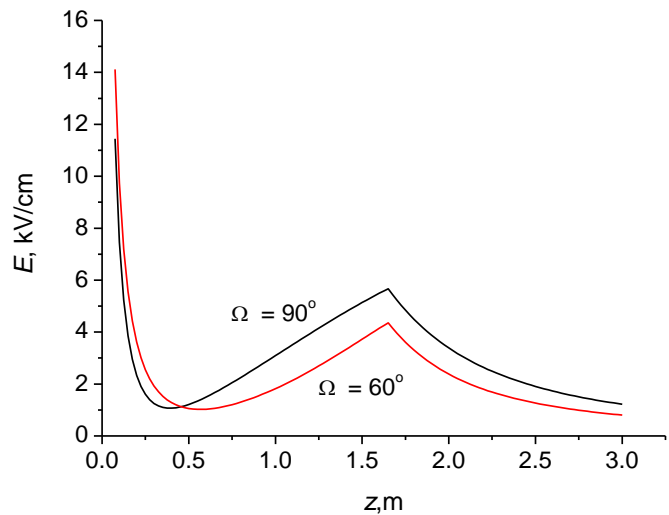

Fig. 2b. Electric field distribution in the streamer zone of a leader: $\rho = \rho_0$.

Consider now the electric field behavior in the case of charges distribution: $\rho \sim 1/r^2$. The expression for the electric field has the form:

$$E = -\frac{\rho_0 R_0^2}{2\varepsilon_0}\{F(z) - F(2H - z)\} + E_{el}, \tag{5}$$

where

$$F(z) = \left(\frac{f(z) - zf'(z)}{z^2} - \frac{f(2H - z) - (2H - z)f'(2H - z)}{(2H - z)^2}\right),$$

$$F(2H - z) = F(z)|_{z=2H-z}$$

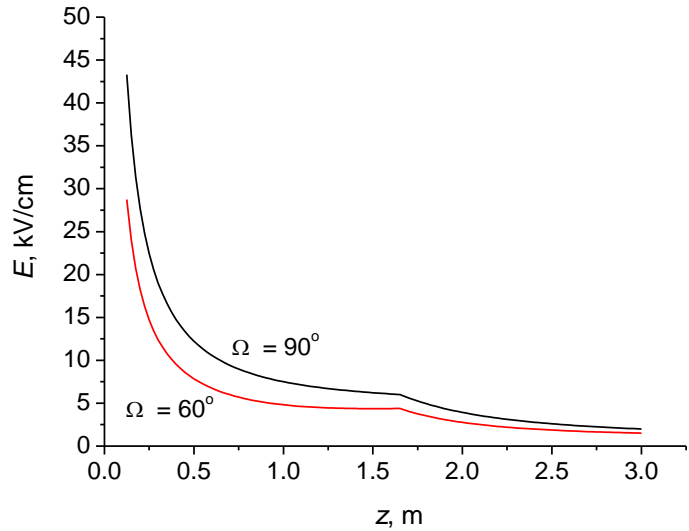

Fig.2c. Electric field distribution in the streamer zone of a leader: $\rho \sim 1/r^2$, $H = 10$ m.

It is seen from the figure 2c that the electric field decreases in a direction away from the top of streamer corona. However for the angles $30° < \Omega < 90°$ the electric field in greater part of streamer corona becomes constant. This agrees with the experimental measurements. This indicates that the charge density distribution in a streamer corona drops with distance by quadratic law: $\rho \sim 1/r^2$.

## 3. TOTAL CHARGE IN A STREAMER ZONE OF LEADER

Determination of the total value of volume charge of streamer corona or streamer zone of a leader is of practical interest. Calculating the integral $Q_\Sigma = \int \rho(r)dV$ for different laws of charges distribution, we obtain:

$$Q_\Sigma = \frac{2\pi}{3}\rho_0 R_0^3(1-\cos(\Omega/2)), \rho = const;$$

$$Q_\Sigma = \pi\rho_0 R_0^3(1-\cos(\Omega/2)), \rho \propto 1/r; \qquad (6)$$

$$Q_\Sigma = 2\pi\rho_0 R_0^3(1-\cos(\Omega/2)), \rho \propto 1/r^2.$$

The parameters $R_0$ and $\Omega$ are known from the experimental measurements. So, in the rod-plane gap with length $H = 20$ m at the duration of front of applied voltage $\tau_f = 300$ µs the streamer zone length is equal to $l_{str} = 1.65$ m and the angle of streamer zone $\Omega = 70°$ [2].

The charge density distribution $\rho_0$ may be also determined if the electric field intensity on the streamer corona boundary $E_{str}$ is known. Actually, using the Gauss theorem the total charge of streamer corona may be represented in the form [13]:

$$Q_\Sigma = \varepsilon_0 \oint E_{str} dS = 2\pi\varepsilon_0 E_{str} l_{str}^2 (1-\cos(\Omega/2)), \qquad (7)$$

where $l_{str} = R_0$.

Equating the expressions (6) and (7), we find that the charge density on the streamer corona boundary is

$$\rho_0 = \varepsilon_0 E_{str}/l_{str} = 2.7 \cdot 10^{-12} C/m^3.$$

Evaluate now the charge density $\rho_{l.h.}$ and their concentration $n_e$ in a leader head supposing that the charge density distribution $\rho \sim 1/r^2$ extends up to the leader head. For the radius of a leader head $r_{l.h.} = 3$ mm and $R_0 = 1.65$ m we obtain: $\rho_{l.h.} = \rho_{l.h.} R_0^2/r^2_{l.h.} \cong 0.8 \cdot 10^{-6}$ C/cm$^3$. The charge concentration in a leader head is equal then $n_e \cong 10^{13}$ cm$^{-3}$, that is in agreement with the known data.

4. **Space-time distribution of electric field intensity created by long spark discharge**

The change of the electric field intensity in the discharge gap is caused by the time dependence of potential of electrodes and dynamics of space charges of the streamer zone and leader channel.

The waveform of electric field intensity obtained using Pockel's device [2, 3] in rod-plane discharge gap with length of $H = 6$ m at positive applied voltage is presented in Fig. 3.

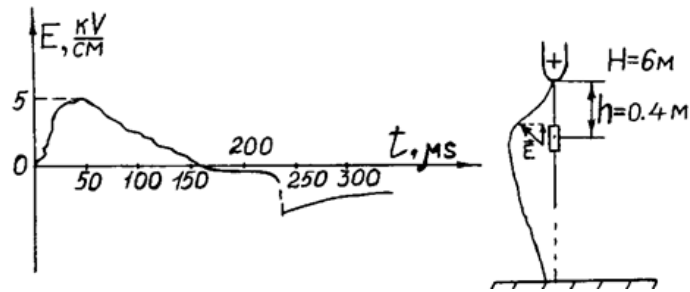

Fig. 3. Measured electrostatic field waveform nearby the positive leader channel in air gap [3].

One can distinguish two stages in the waveforms which correspond to different physical processes. First stage corresponds to the leader phase of a discharge, second stage - to the neutralization of the space charges introduced into the gap (return stroke phase). The duration of leader phase is determined by the leader propagation velocity and it increases with the growth of gap length. At breakdown moment the amplitude of the field increases sharply up to the value of 5 kV/cm during the time less than 1 μs. Electric field decreases smoothly after breakdown.

At negative polarity of the applied voltage the electric field intensity has negative polarity at the leader phase and positive polarity at the return stroke phase (Fig. 4).

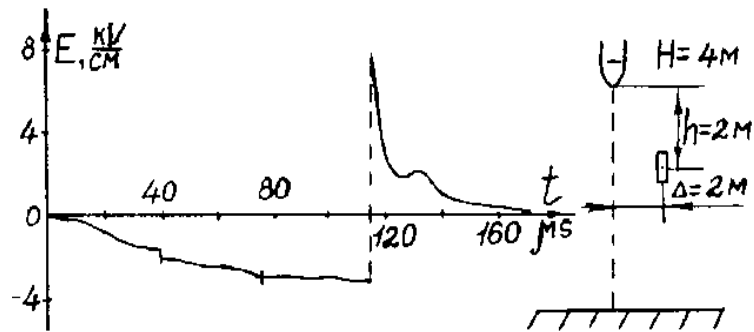

**Fig. 4.** Measured electrostatic field waveform nearby the negative leader channel in air gap with length $H = 4$ m [3].

It is seen that the sharp changes of electric field take place which are caused by the step propagation of negative leader. The time intervals between these fast changes of electric field correspond to the time of the step formation and they compose about 40 μs. Note that the neutralization time of space charges is essentially less than for positive polarity discharge. It takes place also for lightning discharges.

As follows from the measured waveforms, the sub-microsecond changes of field are caused both by the formation of steps in a leader phase and formation of the final stage of breakdown.

In Fig. 5 the electric field behavior in the air discharge gap is shown. The distance between the rod electrode end and sensor was $h = 2$ m.

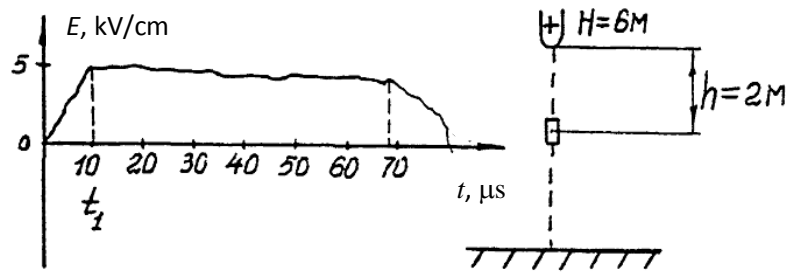

Fig. 5. Electrostatic field waveform inside the streamer zone of the positive leader in rod-plane air gap with length H = 6 m [3].

It is seen that the value of the electric field inside the streamer zone is kept almost constant inside the streamer zone of a positive leader.

## 5. FINAL JUMP PHASE MODEL

Consider the propagation of leader channel/streamer zone system taking into account the influence of the "image" charges (Fig. 6). The streamers are assumed to be propagated up to distance where the electric field is equal to 5 kV/cm.

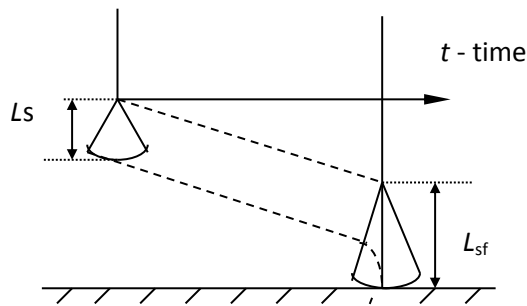

Fig. 6. Leader channel/streamer system evolution in long air gap.

In Fig. 7 the simulation result of the evolution of the streamer-leader system in time is presented. It is shown that the sharp increase in the streamer zone length occurs at the final jump phase.

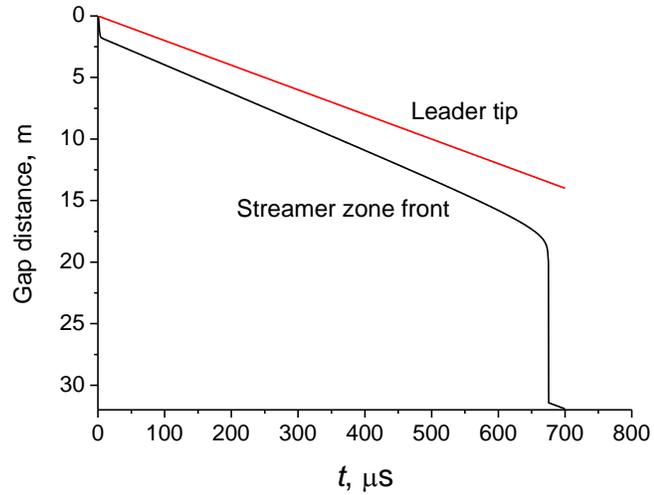

Fig. 7. The evolution of the leader tip and the streamer zone along the gap.

In Fig. 8 the streak photo of a positive leader discharge development in a rod-plane gap with length $H = 10$ m is presented.

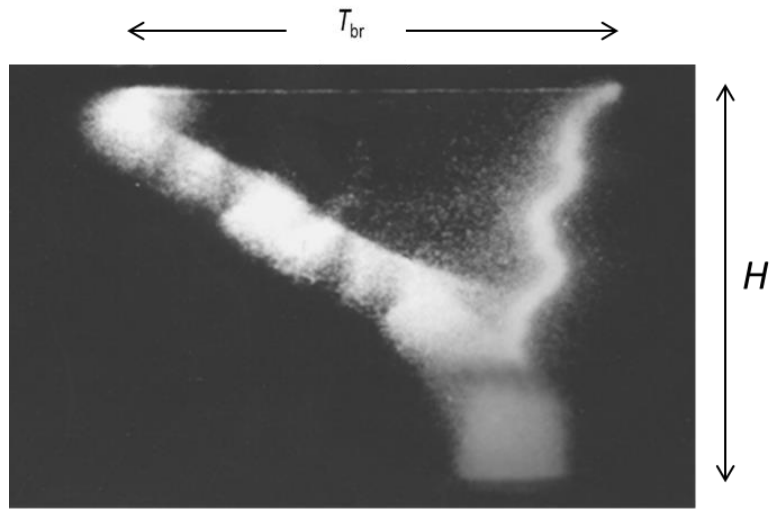

Fig. 8. Streak photograph of a positive leader/streamer developing in a long rod-plane gap [2, 3]. The gap length $H = 10$ m, the leader developing time up to breakdown $T_{br} = 410$ μs.

## 6. DISCUSSION AND CONCLUSIONS

Thus, the potential and electric field behavior in a streamer corona depends on the angle $\Omega$ in the top of streamer corona. The decrease of the angle $\Omega$ leads to the decrease of electric field intensity. It follows from the measured distribution of the electric field inside the streamer corona, that for real values of the angle $\Omega$ the volume charge density distribution $\rho$ in the streamer corona and streamer zone of a leader in long air gaps drops in inverse proportions to the square of distance.

It is known that the final jump phase of a leader determines the striking distance to the objects. Thus the results obtained may be useful in the determination of the striking distance of lightning to earthed structures [15-17].

In recent years, much attention has been paid to the study of energetic radiation pulses at the initial stage of lightning discharge, which precedes the step leader and the return stroke [18]. X-rays with energies extending up to a few hundred keV were also detected during the stepped-leader phase of negative natural lightning strokes [19].

Recently, observations of X-rays from laboratory sparks created in the air at atmospheric pressure by applying an impulse voltage were reported [20]. It was found that the X-rays apparently occurred before the complete breakdown, during the final jump process.